PAPER

# Tracing the chemical footprint of shocks in AGN-host and starburst galaxies with ALMA multi-line molecular studies

Ko-Yun Huang *[a] and Serena Viti [ab]

Multi-line molecular observations are an ideal tool for a systematic study of the physico-chemical processes in the Interstellar Medium (ISM), given the wide range of critical densities associated with different molecules and their transitions, and the dependencies of chemical reactions on the energy budget of the system. Recently high spatial resolution of typical shock tracers – SiO, HNCO, and $CH_3OH$ – have been studied in the potentially shocked regions in two nearby galaxies: NGC 1068 (an AGN-host galaxy) (Huang *et al.*, *Astron. Astrophys.*, 2022, **666**, A102; Huang *et al.*, in prep.) and NGC 253 (a starburst galaxy) (K.-Y. Huang *et al.*, *arXiv*, 2023, preprint, arXiv:2303.12685, DOI: 10.48550/arXiv.2303.12685). This paper is dedicated to the comparative study of these two distinctively different galaxies, with the aim of determining the differences in their energetics and understanding large-scale shocks in different types of galaxies.

## 1 Introduction

The evolution of galaxies is impacted by many key physical and chemical processes in the interstellar medium (ISM) that are often associated with star-formation, Active Galactic Nuclei (AGN), large-scale outflows, and shocks. In this context, nearby galaxies that host starburst and AGN-dominated environments are prime laboratories for the investigation of these feedback mechanisms and their impact on the ISM.

An AGN is considered a manifestation of a mass-accreting supermassive black hole (SMBH) at the center of galaxies, which convert gravitational potential energy and kinetic energy into radiation and powers the jet and the associated multi-phase mass outflow. As AGNs are often identified by prominent point sources in the K band with an associated hard X-ray peak and a bright and compact radio continuum source, the gas chemistry in the nuclear regions of AGN-host galaxies

[a]Leiden Observatory, Leiden University, PO Box 9513, NL-2300 RA Leiden, The Netherlands. E-mail: kyhuang@strw.leidenuniv.nl

[b]Department of Physics and Astronomy, University College London, Gower Street, London WC1E 6BT, UK

is often considered to be X-ray dominated regions (XDRs). Moreover, AGN-dominated galaxies are often thought to have an enhanced cosmic ray ionization rate (CRIR) and radiation field compared to normal spiral galaxies such as the Milky Way. NGC 1068 is a nearby ($D = 14$ Mpc,[1] $1'' \sim 70$ pc) Seyfert 2 galaxy and is considered to be the archetype of a composite AGN-starburst system. The proximity of this galaxy makes it an ideal target for studying in detail the feedback of the AGN activity at its central $r \sim 200$ pc circumnuclear disc (CND). It also allows the simultaneous study of the starburst (SB) activities at the outer $r \sim 1$–1.5 kpc SB ring with the SB ring being spatially resolved from the AGN feedback from the central region of the galaxy. The molecular gas in the CND of NGC 1068 was shown to be outflowing[2] in multi-line CO, HCN, and $HCO^+$ observations.[2,3] The deprojected outflow velocity of NGC 1068 at the CND region was estimated ranging between 85–200 km $s^{-1}$ with the increase in radial distance that runs between $r \sim 50$–200 pc.[3] The outflowing molecular gas in the CND of NGC 1068 is a manifestation of ongoing AGN feedback,[3] for the outflow is likely launched by the interaction between the molecular gas in the CND and both AGN ionized wind and the radio jet plasma.[3] This interaction has produced a large-scale molecular shock on spatial scales of up to $r = 400$ pc from the AGN and likely triggers rich shock chemistry signatures across the CND.[4]

Starburst galaxies are extremely luminous systems that are powered by bursts of massive star formation. Strong stellar feedback induced by high star-forming rates (SFR) can trigger multi-phase outflows that involve ionized, neutral, and molecular gas. As starburst activities inject a significant amount of energy into the ambient environment, starburst galaxies are also important targets to study the feedback mechanisms in the ISM. NGC 253 is a barred spiral galaxy that is almost edge-on with an inclination of 76°,[5] and is one of the nearest starburst systems ($D \sim 3.5 \pm 0.2$ Mpc (ref. [6])). NGC 253 is considered a prototype of nuclear starburst with an SFR of $\sim$2 $M_\odot$ $yr^{-1}$ coming from its central molecular zone (CMZ),[7,8] which is half of its global SF activity. The CMZ of NGC 253 spans about $300 \times 100$ pc across.[9] A large-scale outflow in NGC 253 has been revealed by multi-wavelength observations: in X-rays,[10,11] H$\alpha$,[12] molecular emission,[13–16] and dust.[17] This large-scale outflow is thought to be driven by the galaxy's starburst activity,[18] for there are no signs of AGN influence[19,20] despite there being a bright compact radio source associated with the nucleus of the galaxy.[21] The deprojected outflow velocity of NGC 253 was estimated as ranging between 110–360 km $s^{-1}$ which increases with distance above the plane up to $r \sim 1$ kpc.[12] The presence of shocks in NGC 253 has been suggested by the detection of two molecular shock tracers, HNCO and SiO,[22,23] as well as the detection of Class I $CH_3OH$ masers,[24] and the enhanced fractional abundances of $CO_2$.[25]

Both silicon monoxide, SiO, and isocyanic acid, HNCO, are well-known shock tracers,[26–31] and have been observed in nearby galaxies.[22,23,32–39] The simultaneous detection of HNCO and SiO has been found to be very useful for the characterization of different types of shocks (*e.g.* fast *versus* slow shocks) in NGC 1068,[38,39] NGC 1097,[37] in the nearby weakly barred spiral galaxy IC 342 which hosts moderate starburst activities,[32,33] and in NGC 253.[23] A high abundance of silicon in the gas phase can only be explained by significant sputtering from the core of the dust grains by high-velocity ($v_s \geq 50$ km $s^{-1}$) shocks.[38] Once silicon is in the gas phase, it is expected to quickly react with molecular oxygen or a hydroxyl radical to form SiO.[41] An enhanced gas-phase SiO abundance could thus be a sensitive indicator of the heavily shocked regions. The formation of HNCO has been suggested to be mainly on the icy mantles of dust grains,[42] or possibly in the gas phase with subsequent freeze-out onto the dust grains when the temperatures are low.[43] Regardless of how it forms, icy mantles sputtering associated with low-velocity ($v_s \leq 20$ km s$^{-1}$) shocks can lead to an enhanced abundance of HNCO in the gas phase. The good correlation between HNCO and SiO revealed in Galactic dense molecular cores by Zinchenko *et al.*[28] hinted that both species trace shocks, although the absence of HNCO in the higher velocity wings observed in the SiO spectral profile also hinted to the fact that high-velocity shock conditions may suppress the HNCO abundance. A follow-up survey over sources towards the Galactic Center performed by Martín *et al.*[30] reveals that HNCO can however also be heavily destroyed by UV radiation in PDR regions (later also found in NGC 253 in Martín *et al.*[34]). On the other hand, chemical modelling of HNCO and SiO in NGC 1068 performed by Kelly *et al.*[38] confirmed that the HNCO abundance can be suppressed in high-velocity shocks due to the destruction of its precursor, the molecule NO. In fact, Kelly *et al.*[38] showed that HNCO can also be thermally desorbed from the surface of dust grains when the gas and dust remain coupled at higher gas densities ($n_{H_2} \geq 10^4$ cm$^{-3}$). Hence HNCO is a complex molecule to interpret and may not be a unique tracer of shock activity.

In our two recent works, we presented the observations of multiple transitions of HNCO and SiO observed with ALMA towards both the CND of NGC 1068 [39] and the CMZ of NGC 253 (Huang *et al.*,[40]). These two studies have shown clear signatures of shocks in these two different environments under the influence of shocks from potentially multiple origins.

In this paper we revisit these observational results and present an astro-chemical modeling study with the aim of comparing these two galaxies in HNCO and SiO. For completeness, we shall also include a brief comparison with the results from our most recent work (Huang *et al.*, in prep.) where we explore the shock properties as traced by $CH_3OH$ in the CND of NGC 1068. The aim will be (1) to determine whether these three species (HNCO, SiO, and $CH_3OH$) are uniquely sensitive to certain type(s) of shocks in the two galaxies in consideration; (2) to see if we can get a handle on determining the shock origins in these regions; and finally (3) to explore the possibility of differentiating the energetics of these shocks (*e.g.* AGN-driven or starburst-driven) in different types of galaxies.

In Section 2 we briefly summarize the observations of HNCO and SiO in NGC 253 and NGC 1068. In Section 3 we describe our chemical modeling analysis of HNCO and SiO. Section 4 covers a brief discussion of $CH_3OH$. In Section 5 we discuss the shock properties that each species is tracing, their shock history and structure of the shocks. We also explore the potential in differentiating the shock chemistry in the AGN-dominated *versus* starburst environments, on the basis of these two galaxies – NGC 1068 and NGC 253. Our conclusions are summarized in Section 6.

## 2 Summary of previous observational results

In Huang *et al.* 2022[39] we examined the potential shock signatures traced by HNCO and SiO in the CND of NGC 1068 at a spatial resolution of ~56 pc, which is comparable to the size of a giant molecular cloud (GMC), using ALMA observations. Fig. 1 shows some of the line intensity maps from Huang *et al.*[39] from HNCO and SiO, and their relevant layout with the outflow and the host galaxy itself shown in the dust continuum map adapted from García-Burillo *et al.*[2]

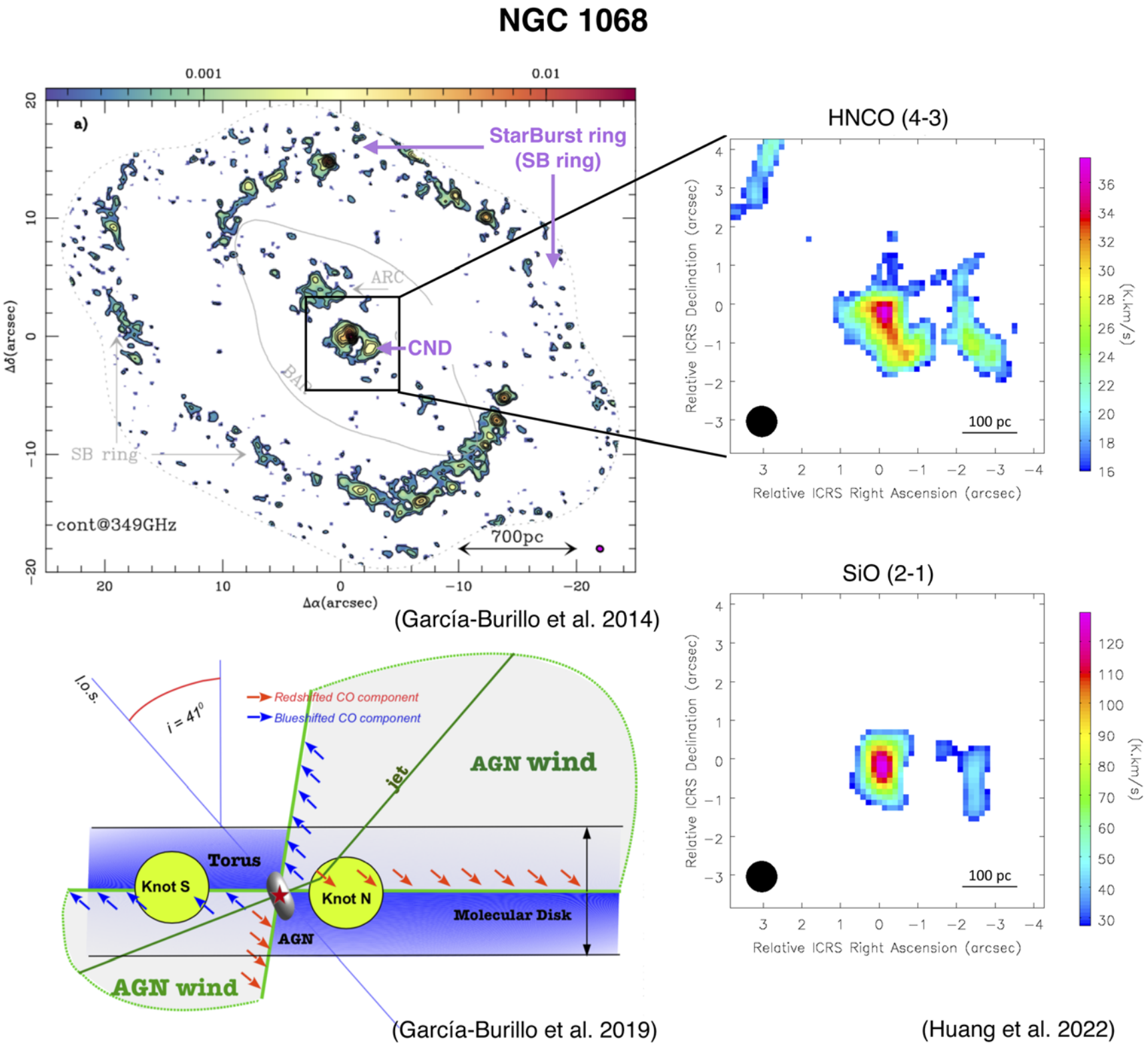

**Fig. 1** The two example intensity maps, HNCO (4-3) and SiO (2-1), from the CND of NGC 1068 (two maps on the right, adapted from Huang *et al.*[39]) relative to the layout of the galaxy. *Top left:* dust continuum map of NGC 1068 including the SB ring, adapted from García-Burillo *et al.*;[2] *bottom left*: a schematic adapted from García-Burillo *et al.*,[3] illustrating the relative layout of the AGN wind, the molecular disk/torus, and the line of sight of the observer for NGC 1068.

In Huang *et al.*,[40] the same pair of shock tracers, HNCO and SiO, were used to study the CMZ of NGC 253 at higher spatial resolution of ∼28 pc using ALMA observations as part of the ALCHEMI large program. Fig. 2 shows examples of the line intensity maps from Huang *et al.*[40] of HNCO and SiO, and their relevant layout with the outflow and the host galaxy itself with maps adapted from Bolatto *et al.*[14]

We list below the main observational results and the remaining open questions from these two previous studies that are relevant to the current work:

(1) From both studies it was confirmed that HNCO and SiO are consistently tracing very different gas components. HNCO tends to trace cooler and denser gas components than those traced by SiO. The high gas temperatures (few hundreds K) traced by SiO signal the evidence of strong shock heating.

(2) In NGC 1068, the discussion of the chemical origin(s) of the two species in Huang *et al.* 2022[39] is based on the prior chemical modeling performed by Kelly

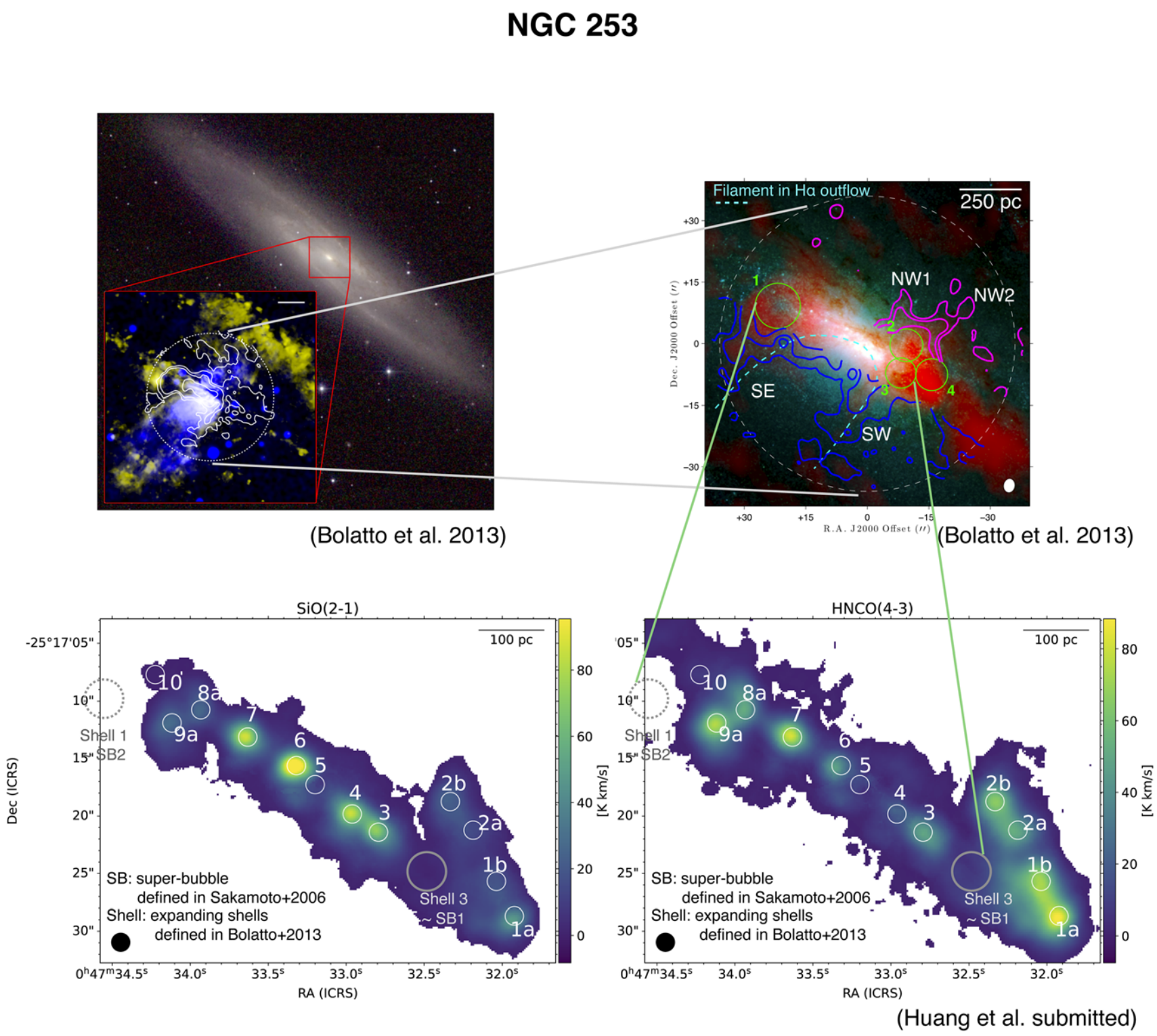

**Fig. 2** The two example intensity maps, HNCO (4-3) and SiO (2-1), from the CMZ of NGC 253 (two maps on the bottom, adapted from Huang *et al.*[40]) relative to the layout of the galaxy. *Top panels*: maps illustrating the multi-phase galactic wind in NGC 253, adapted from Bolatto *et al.*[14] and reference therein.

*et al.* 2017[38] with only low CRIR assumed for the CND ($\zeta = \zeta_0$). In the latter, SiO was clearly associated with strong shocks, yet the chemical origin of HNCO could either be weak shocks or thermal sublimation in denser gas ($n_{H_2} \geq 10^4$ cm$^{-3}$) environment. The chemical modeling performed by Huang *et al.*[40] strongly supports the fast-shock origin of SiO and slow-shock origin of HNCO in the cases tailored for the high cosmic-ray ionization rate (CRIR) environment ($\zeta = 10^{3-5}\zeta_0$, $\zeta_0 = 1.3 \times 10^{-17}$ s$^{-1}$ is the standard galactic CRIR) of NGC 253. Huang *et al.*[40] shows higher CRIR not only affects both HNCO and SiO abundances in shock chemistry, but makes the thermal sublimation an unviable route in enhancing HNCO abundance.

(3) With well-constrained gas density and gas temperature inferred from the RADEX modeling coupled with Bayesian inference processes that are informed with observational data, Huang *et al.*[40] have shown it to be possible to further infer the age since the last shock episode with each shock tracer. This is because the shock influence timescale is inversely proportional to the gas density. Gas components showing different gas temperatures suggest that they are at different post-shock cooling stages relative to this shock influence timescale.

(4) It remained unclear, however, whether the weak shock episode traced by HNCO and the strong shock episode evidenced by SiO – both in NGC 1068 and NGC 253 – point to the same occurrence of shock. In other words, it is possible that the two types of shock episodes arise from events that are not associated in causality. However, if one associates the weak shock episode and the strong shock episode to be the same occurrence of shock event, Huang *et al.*[40] showed that there is a global trend in the age of shocks, which increases from the inner to the outer regions of the CMZ of NGC 253.

(5) The chemical origin of the observed HNCO at gas density $n_{H_2} \sim 10^3$ cm$^{-3}$ remains unclear. Neither shock scenarios nor thermal sublimation can account for the HNCO abundance observed at this low gas density in the CMZ of NGC 253.

The summary of key properties concerning these two galaxies that will be used in the current work are listed in Table 1. The listed fractional molecular abundance is derived based on the observational data and the gas property inference *via* radiative transfer modeling presented by Huang *et al.* (2022)[39] and Huang *et al.*[40] We call this observation-based fractional abundance $X_{\text{species,obs}}$. In particular, from the gas property inference performed by both previous studies, both molecular gas density ($n_{H_2}$) and the molecular total column density per species ($N_{\text{HNCO}}$ and $N_{\text{SiO}}$) can be obtained. These gas properties can lead to an estimate of an "observation-based" fractional molecular abundance by the following relation:

$$X_{\text{species,obs}} = \frac{N_{\text{species}}}{n_{H_2}\Delta z_{\text{cloud}}\eta_{\text{ff}}} \sim \frac{N_{\text{species}}}{n_{H_2}\theta_{\text{beam}}\eta_{\text{ff}}} \geq \frac{N_{\text{species}}}{n_{H_2} \times \theta_{\text{beam}} \times 1.0}, \tag{1}$$

where $\Delta z_{\text{cloud}}$ refers to the line-of-sight dimension of the gas component in consideration, $\theta_{\text{beam}}$ stands for the beam size of the observation, and $\eta_{\text{ff}}$ is the beam filling factor which is a parameter illustrating the source size relative to the beam size as defined by $\eta_{\text{ff}} = \dfrac{{\theta_S}^2}{{\theta_{\text{beam}}}^2 + {\theta_S}^2}$. As an approximation, we use the beam size of the observation ($\theta_{\rm beam}$) as the estimate of the line-of-sight dimension of the gas component ($\Delta z_{\rm cloud}$), and in the final inequality we used the 1.0 as the upper limit for the $\eta_{\rm ff}$.

Table 1 List of physical and chemical properties of the CND in NGC 1068 and the CMZ of NGC 253 based on previous observational studies[a]

| | NGC 1068 | | NGC 253 | |
|---|---|---|---|---|
| Property | CND | Ref. | CMZ | Ref. |
| Beam size [pc] | 56 | (a) | 28 | (b) |
| $n_{H_2}$[HNCO] | $10^3$ to $10^6$ | (a) | $10^3$–$10^6$ | (b) |
| $n_{H_2}$[SiO] | $10^3$ to $10^4$ | (a) | $10^2$–$10^4$ | (b) |
| $X_{\text{HNCO,obs}}$ | $\geq(2.71 \times 10^{-12}$ to $1.21 \times 10^{-8})$ | | $\geq(2.65 \times 10^{-12}$ to $6.60 \times 10^{-8})$ | |
| $X_{\text{SiO,obs}}$ | $\geq(1.13 \times 10^{-10})$ | | $\geq(2.47 \times 10^{-8}$ to $1.61 \times 10^{-7})$ | |
| $\zeta$ [$\zeta_0$] | 1.0–10.0 | (c) | $10^3$–$10^5$ | (d, e, f and g) |
| $v_{\text{outflow}}$ [km s$^{-1}$] | 85–200 | (h) | 110–360 | (i) |

[a] References: (a) Huang *et al.* 2022;[39] (b) Huang *et al.*[40] (submitted); (c) Scourfield *et al.* 2020[44] – CRIR inferred from multi-J CS emissions in the CND region of NGC 1068; (d) Holdship *et al.* 2021;[45] (e) Harada *et al.* 2021;[46] (f) Holdship *et al.* 2022;[47] (g) Behrens *et al.* 2022;[48] (h) García-Burillo *et al.* 2019;[3] (i) Westmoquette *et al.* 2011[12] – this estimate is up to radial distance of 1 kpc.

## 3 Modeling shock chemistry

As mentioned earlier, for NGC 1068, Huang *et al.* (2022)[39] did not model the chemistry but based our conclusions on prior modeling work by Kelly *et al.*[38] Here, therefore, we present a chemical modeling analysis that covers the most comprehensive set of physical and chemical conditions which takes both galaxies – NGC 1068 and NGC 253 – into consideration.

The chemical modeling was performed with the open source time dependent gas–grain `UCLCHEM` code[49] (**https://uclchem.github.io**) in order to further explore the chemical origin of the observed HNCO and SiO emissions in these two distinctively different galaxies. `UCLCHEM` is a 3-phase gas–grain chemical modelling code that incorporates user-defined chemical networks to produce chemical abundances along user-defined physics modules that can simulate a variety of physical conditions.

We perform modeling cases that cover the physical conditions measured from the CND of NGC 1068 and the CMZ of NGC 253 – this includes a wide CRIR range ($\zeta = 1.0$–$10^5\zeta_0$, $\zeta_0 = 1.3 \times 10^{-17}$ s$^{-1}$ is the standard galactic CRIR), slow ($v_{\rm s} = 10$ km s$^{-1}$) *versus* fast ($v_{\rm s} = 50$ km s$^{-1}$) shocks, and a wide range of molecular gas densities $n_{\rm H_2}$. The chosen CRIR range is based on the existing CRIR measurements[44–48] in these two extragalactic environments. In particular, in the CND of NGC 1068, the inferred CRIR based on multi-transition CS observations ranges between $\zeta = 1.0$–$10.0\zeta_0$.[44] And in the CMZ of NGC 253, the measured CRIR based on multi-molecule, multi-transition observations is estimated to be $\sim\zeta = 10^{3-5}\zeta_0$,[45–48] which is a few orders of magnitude higher than NGC 1068. A brief summary of the parameter space explored in our chemical modeling is listed in Table 2.

`UCLCHEM v3.1` includes an improved sputtering module in the parameterized C-shock model following Jiménez-Serra *et al.*[29] as well as a 3-phase chemistry where chemistry is computed for the gas phase, the grain surfaces, and the bulk ice. These improvements, compared to the model used in Kelly *et al.*,[38] ensure a better treatment of the sputtering of refractory species, such as Si-bearing species, during the shock process. Aside from these technical differences, the work by Kelly *et al.* 2017[38] only covers the modeling with standard galactic CRIR ($\zeta_0$).

Table 2 The parameter space explored in our chemical modelling. Note that $\zeta_0 = 1.3 \times 10^{-17}$ s$^{-1}$ and $X_{\rm Si,gas,\odot} = 4.07 \times 10^{-5}$

| Variable | Grid |
|---|---|
| Gas density,[a] $n_{\rm H_2}$ [cm$^{-3}$] | [$10^3$, $10^4$, $10^5$, $10^6$] |
| C-shock velocity, $v_{\rm shock}$ [km s$^{-1}$] | [10.0, 50.0] |
| CRIR $\zeta$ [$\zeta_0$] | [1.0, 10.0, $10^3$, $10^5$] |
| Physical model | Shock or non-shocked scenario |

[a] Gas density refers to the pre-shock gas density in the shock scenario, and the initial gas density in the non-shocked scenario.

In Fig. 3–6 we show the results of our modeling for four gas density setups ($n_{H_2}$ = 103 to $n_{H_2}$ =106 cm$^{-3}$, ordered from Fig. 3 to 6) with two shock velocities ($v_s$ = 10 km s$^{-1}$ and $v_s$ = 50 km s$^{-1}$) to represent the slow and fast shock scenarios, and these two shock scenarios are shown in the left (slow shock) and right (fast shock) panels respectively. In each row we show the modeling results using different CRIR from the top row (lowest CRIR) to bottom row (highest CRIR). In particular, in each figure (*e.g.* Fig. 3) the top two rows are models at lower CRIR for the CND of NGC 1068, and the bottom two rows are models at higher CRIR for the CMZ of NGC 253. Each Figure is associated with one specific gas density *e.g.* Fig. 3 at a gas density $n_{H_2} = 10^3$ cm$^{-3}$. In each figure we also overlay the minimum fractional abundances as constrained by the observations in NGC 1068 and NGC 253 respectively, $X_{\text{species,obs}}$, which was explained earlier in Section 2 with the corresponding gas density and CRIR.

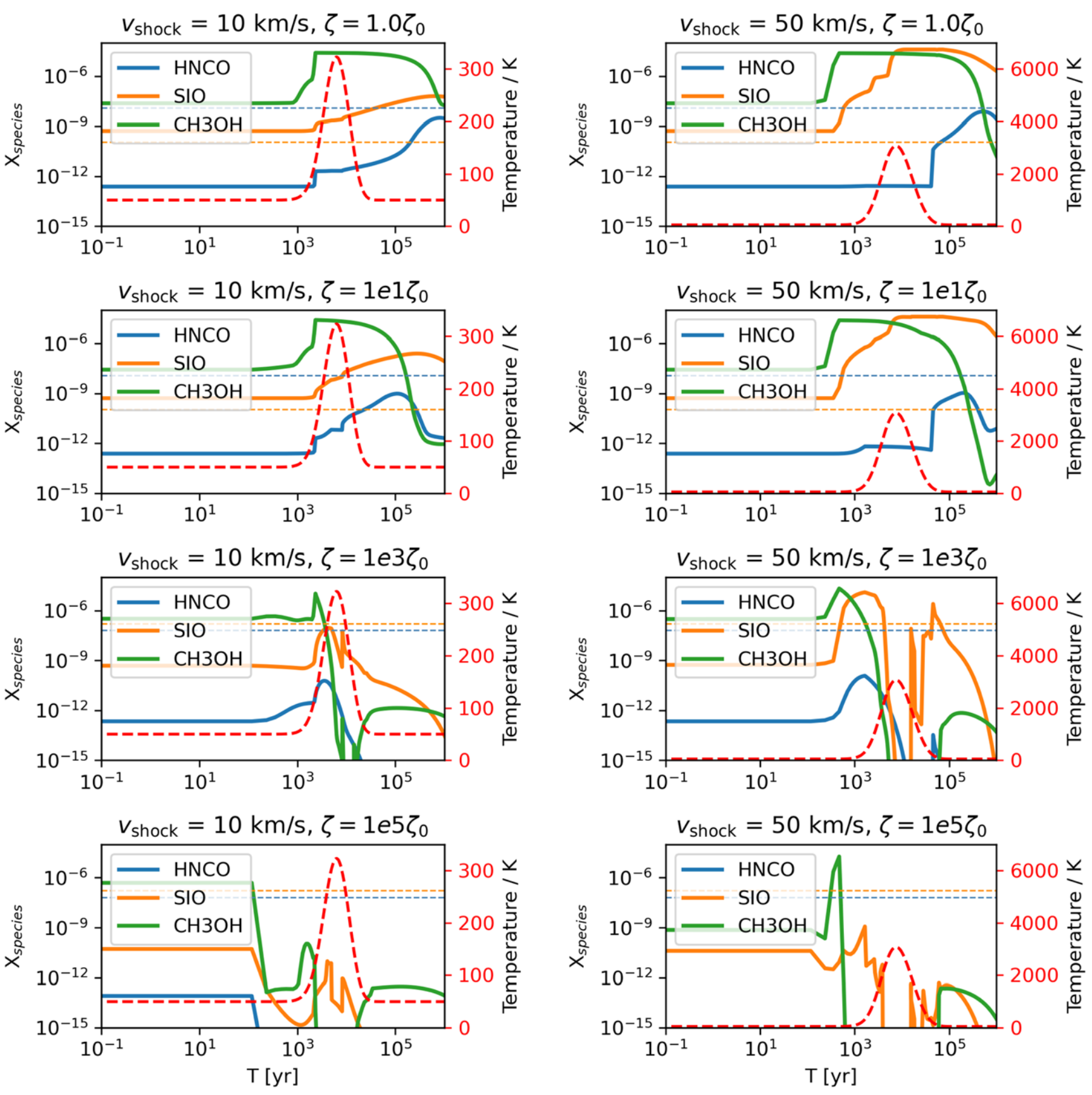

Fig. 3 Chemical abundances as a function of time for slow shock ($v_s$ = 10 km s$^{-1}$) models (left panel) and fast shock ($v_s$ = 50 km s$^{-1}$) models (right panel). The pre-shock gas density in the shock models and the gas density in non-shock models is 10$^3$ cm$^{-3}$. The red dashed curve represents the temperature profile, with the temperature scale on the vertical axis on the right, also in red. For the shock models, within each panel we present from top to bottom: [CRIR ($\zeta$) = $\zeta_0$], [CRIR ($\zeta$) = 10$\zeta_0$], [CRIR ($\zeta$) = 10$^3\zeta_0$], and [CRIR = 10$^5\zeta_0$]. The dashed, colored horizontal lines indicate the lower limit of the species fractional abundances "measured" from our RADEX-Bayesian inference based on observational data and with an assumed hydrogen column density, for HNCO (blue) and SiO (orange) respectively. The fractional abundance values used are the minimum derived values among the studied GMC-sized regions in the two galaxies. In the top 4 figures the "measured" fractional abundances are cited from observations towards the CND of NGC 1068 which was found of lower CRIR with corresponding colors for each species when observed cases are available. On the other hand, the "measured" fractional abundances in the bottom 4 figures are cited from observations towards the CMZ of NGC 253, as found with higher CRIR.

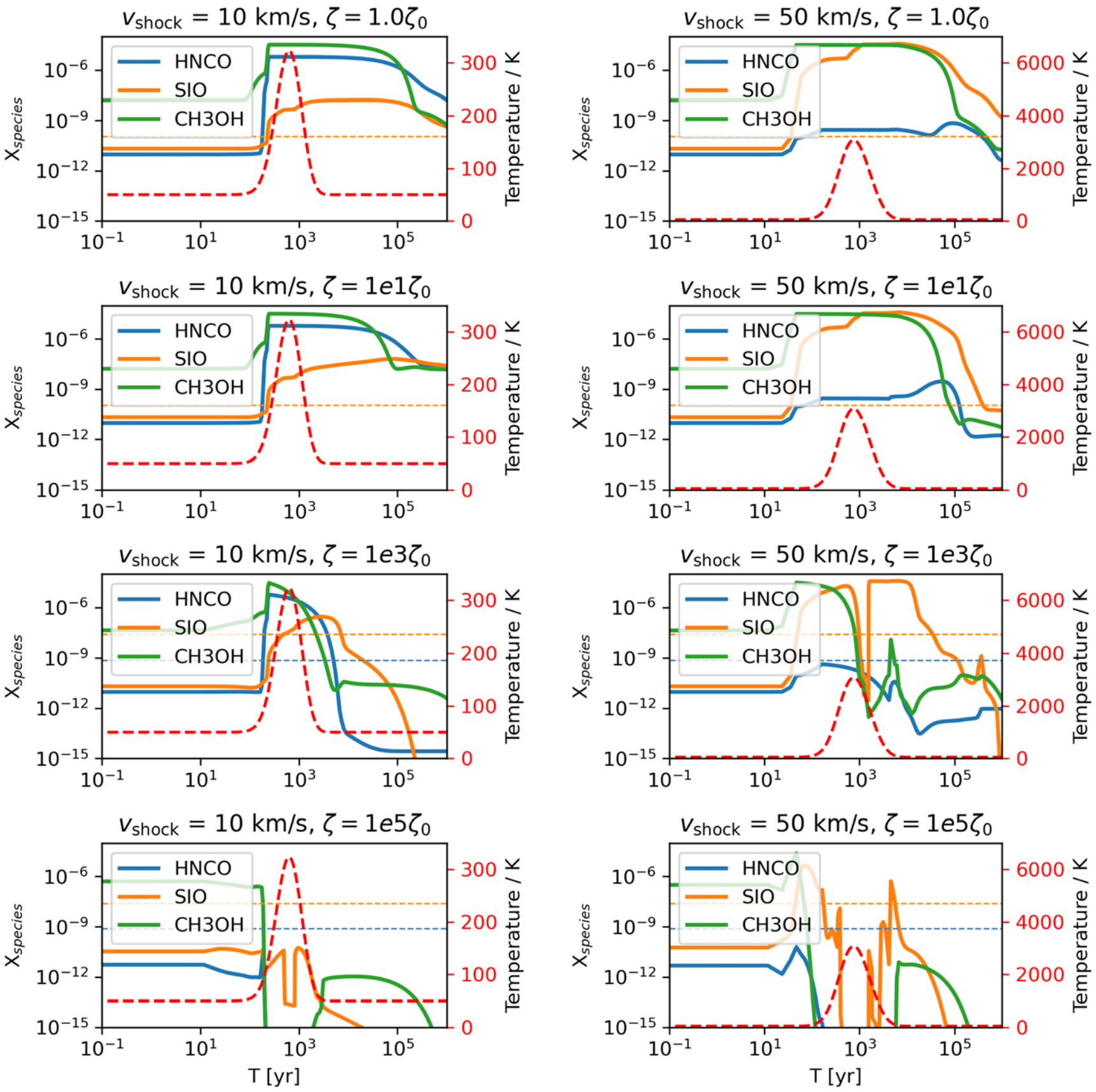

Fig. 4 As in Fig. 3 but for a pre-shock gas density of $10^4$ cm$^{-3}$.

First of all for the SiO abundance (orange solid curves), it is clear that fast-shock chemistry dominates over slow-shock chemistry. While slow shocks can also enhance the SiO abundance, the enhancement is still more than 2 orders of magnitude lower than the enhancement arising from fast shocks. The case for the HNCO abundance (blue solid curves) is reversed, where its enhancement is dominated by slow-shock chemistry. The only exception is in the least dense gas environment ($n_{H_2} = 10^3$ cm$^{-3}$, Fig. 3), where the enhancement of HNCO in none of the cases is sufficient to reach the minimum observed, $X_{\text{HNCO,obs}}$.

In general, high CRIR suppresses the abundances of all shock tracers of interest, especially the highest CRIR tested, $\zeta = 10^5\zeta_0$. In the denser gas environments ($n_{H_2} = 10^{5-6}$ cm$^{-3}$) such suppression seems less severe.

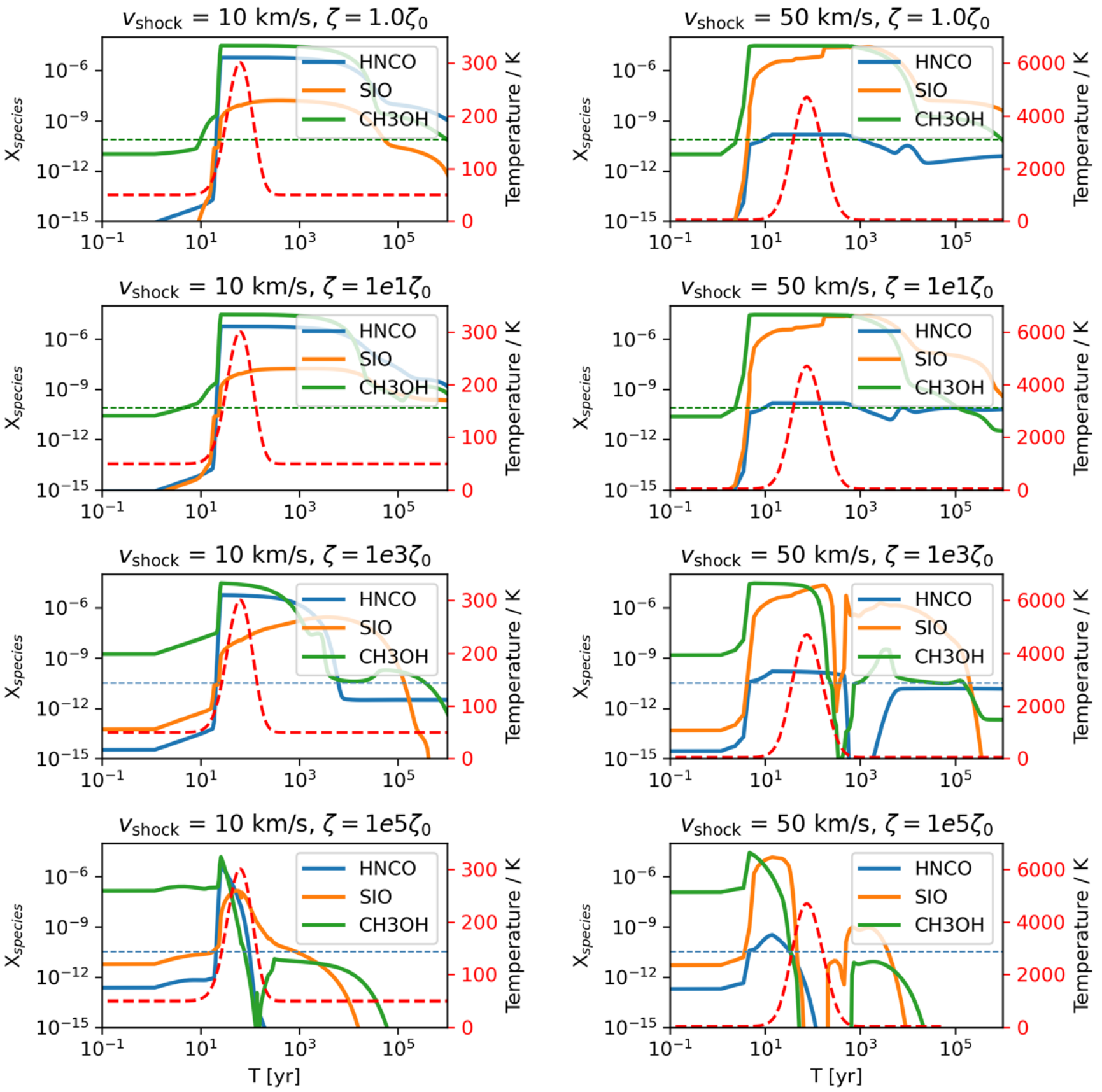

Fig. 5 As in Fig. 3 but for a pre-shock gas density of $10^5$ cm$^{-3}$. The dashed, green horizontal lines in the top 4 figures indicate the lower limit of the species fractional abundances of methanol "measured" from our RADEX-Bayesian inference based on observational data and with an assumed hydrogen column density for the CND of NGC 1068.

# 4 Methanol in the CND of NGC 1068

Methanol ($CH_3OH$) can be formed efficiently in cold environments (12–20 K) through repeated hydrogenation of CO on interstellar ices[50–54] as well as *via* the radical-molecule H-atom abstraction reaction.[55] Sources of energy such as shocks can remove $CH_3OH$ from the grain surface and enhance the gas-phase $CH_3OH$ either through a sputtering process or sublimation. Although gas-phase formation routes of $CH_3OH$ have also been proposed, it cannot account for the observed $CH_3OH$ in general.[56,57] $CH_3OH$ has been therefore also proposed as a good candidate tracer of shocks in the ISM, although it has also been detected in cold and quiescent environments such as pre-stellar cores[58,59] where alternative non-thermal desorption processes aside from shocks may be needed.

In Fig. 3–6 we also display the $CH_3OH$ fractional abundances (in green solid curves) over slow *versus* fast shock chemistry. It is clear that both slow shock and fast shock are able to enhance the $CH_3OH$ abundance to comparable levels. This is consistent with prior modeling work,[49] where $CH_3OH$ was shown enhanced for a large span of shock velocities. One of the key pieces of information to differentiate the shock origin – fast or slow shock – of the observed $CH_3OH$ is the gas temperature. In the slow-shock scenario the gas temperature can only reach a few hundred Kelvin, while in the fast-shock scenario the gas temperature can reach up to a few thousand Kelvin. Note that, however, a measured low gas temperature does not necessarily mean only slow-shock origin, for it could arise from a gas component that experienced full post-shock cooling down from even a few thousand Kelvin.

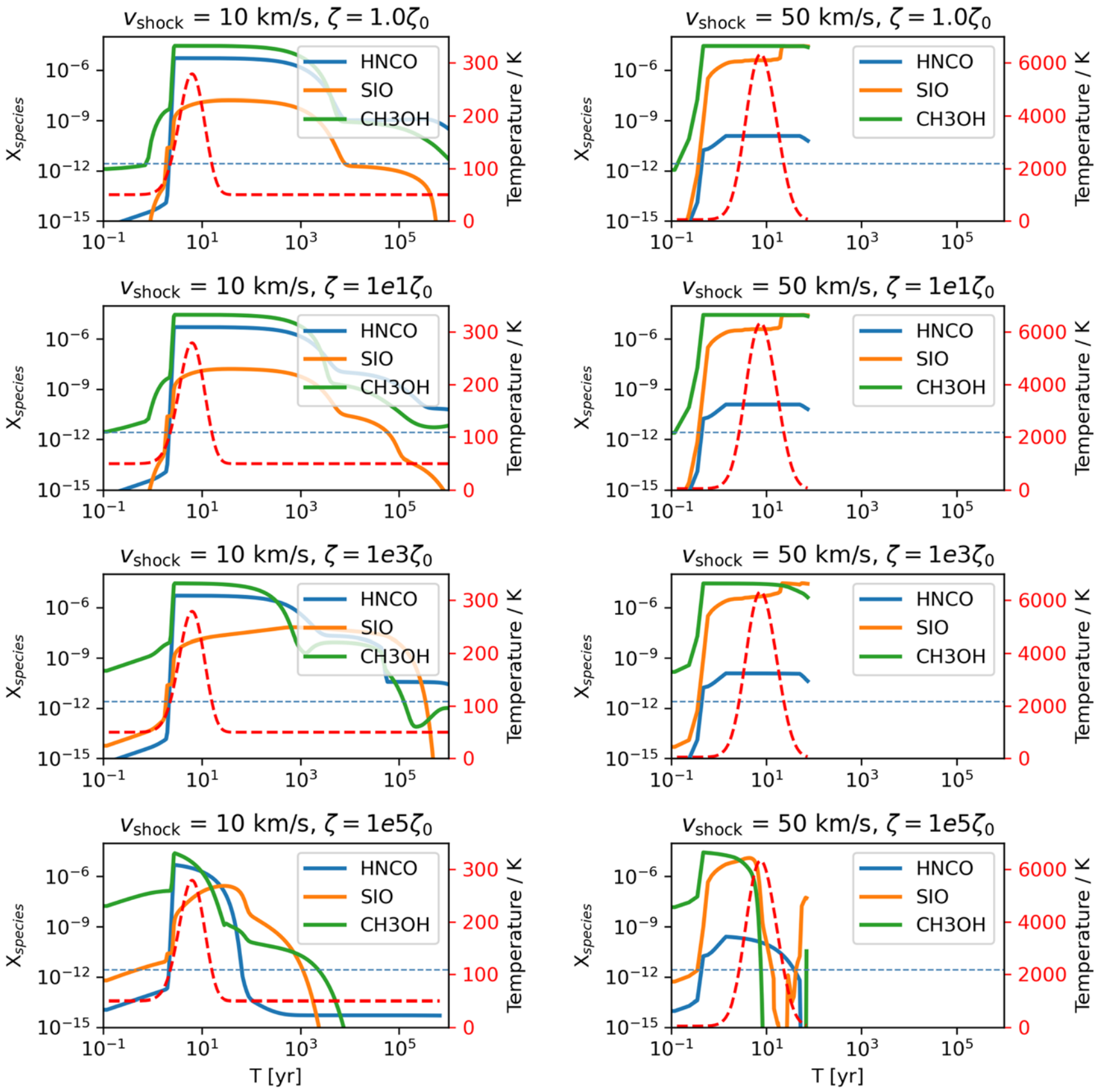

Fig. 6 As in Fig. 3 but for a pre-shock gas density of $10^6$ cm$^{-3}$.

Such degeneracy will be an interesting piece of knowledge when one wants to reconstruct the shock structure and shock history in the target field. As pointed out by Huang *et al.*,[40] it remains unclear whether the fast shock episode traced by SiO and the slow shock episode traced by HNCO are caused by the same shock occurrence. It is possible that they arise from the same shock events but belong to different parts of the gas structure, or they could be triggered by sporadic events that are not linked at all. As $CH_3OH$ is enhanced to comparable levels in both fast shock and slow shock, how well matched in its spatial distribution and the gas properties it traces compared to the other shock-tracing species (*e.g.* HNCO and SiO) will be critical to the overall understanding of the shock structure and history reconstruction.

For instance, our preliminary analysis of methanol data in the CND in NGC 1068 from Huang *et al.*, in prep. indicates that $CH_3OH$ is tracing similarly high-density and cold ($T$ < 50 K) gas components as HNCO in the CND regions. In terms of the chemical origin of the observed $CH_3OH$, it could arise from the post-shock cooling condition gas under the influence of both fast and slow shocks. One should therefore be careful in using these three species at once to portrait the shock structure in the field. The degeneracy and the great complexity are inevitable when we include more and more molecular tracers in consideration, but we want to emphasize that with careful analysis and interpretation it will at the same time provide enormous potential in fully reconstructing the shock structure and history with well-covered parameter space measured from observations.

# 5 Discussion

### 5.1 Spatial distribution of shocked gas in the two galaxies

Although observed with lower spatial resolution (∼56 pc), the HNCO emission tends to be more spatially extended than SiO in the CND of NGC 1068 (*e.g.* Fig. 1), as pointed out by Huang *et al.*[39] It is different for the case of CMZ in NGC 253, where HNCO and SiO emission seems to be co-spatial in most transitions, see Fig. 2 for example.

There are a few possible reasons concerning such differences in the two systems:

• The spatial offset between HNCO and SiO identified in the CND of NGC 1068 (but not in the CMZ of NGC 253) could arise from the difference in the sensitivity levels of the two sets of observations. In other words it is possible that SiO in NGC 1068 is also extended throughout the CND rather than just localized in the East and West spots of the CND.

• As NGC 253 is more edge-on, it is possible that the difference is due to a difference in viewing angle.

• The less extended SiO emission in the CND of NGC 1068 could instead arise from chemical differential across the CND of NGC 1068. It is possible that SiO is highlighting the strongly shocked gas layer(s) and HNCO is mapping out more extended and weakly shocked region, potentially correlated to the geometry of the AGN-driven, large-scale outflow. Meanwhile the shocks induced in the CMZ of NGC 253 may originate from the mini-starburst events in each GMC, rather than from an ordered, large-scale outflow.

### 5.2 Weak shock tracers – degeneracy with non-shock chemistry

As we mentioned in earlier sections (Section 1 and Section 4), both HNCO and $CH_3OH$ can also be enhanced *via* thermal sublimation of the ices induced by events other than shocks. In Fig. 7 we show a chemical model for a gas density $n_{H_2} = 10^5\ cm^{-3}$ in its chemical evolution of the gas being warmed by the presence of star forming processes (including outflows), and/or X-ray or cosmic rays instead of being shocked. This shows that both HNCO and $CH_3OH$ abundances can be enhanced in a warm-gas environment without shocks particularly at low CRIR cases ($\zeta = 1.0$–$10.0\zeta_0$, top two panels). This degeneracy (shock chemistry or thermal sublimation) will impact more the interpretation of the CND in NGC 1068, where lower (than in NGC 253) CRIR values were reported.[44]

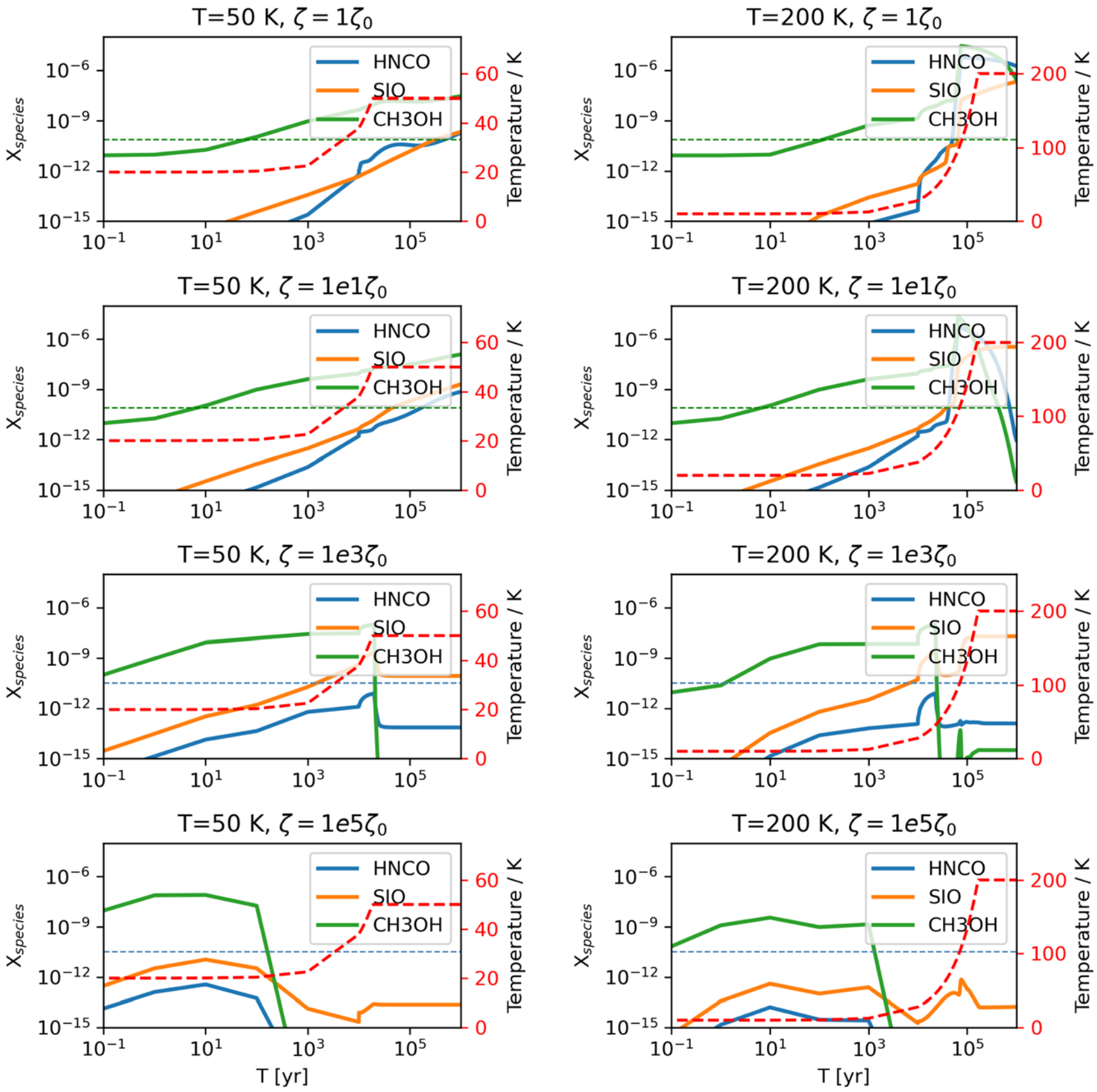

Fig. 7 As in Fig. 3 but for the non-shock models with gas density $10^5$ cm$^{-3}$. The dashed, green horizontal lines in the top 4 figures indicate the lower limit of the species fractional abundances of methanol "measured" from our RADEX-Bayesian inference based on observational data and with an assumed hydrogen column density for the CND of NGC 1068.

### 5.3 Shock history reconstruction in both galaxies and the shock origin

In NGC 253 it remains unclear whether the fast shock episodes traced by SiO are associated with the slow shock episodes traced by HNCO; however if one associates the shock timescales traced by these two species there seems to be a trend of increasing shock ages from the inner to the outer region of the CMZ. Such trend could be related to the large-scale outflow that stems from the nuclear region due to starburst activities. On the other hand, it remains possible that the shock episodes traced by HNCO and SiO were triggered by sporadic, localized events which are only sensitive to their neighboring environment.

In the case of NGC 1068, however, due to the poorly constrained gas temperature, it is still difficult to gauge such shock timescales and make comparison across the CND region. Future observations of more transitions of both species are needed to properly constrain the gas temperature.

### 5.4 Shock chemistry in the two environments – NGC 1068 and NGC 253

**5.4.1 Velocities of AGN-driven *vs.* starburst-driven outflows.** As it is also often thought that AGN-driven outflows can be more powerful than those starburst-driven, we also inspected the outflow velocities from the two systems reported in the literature as listed in Table 1. It seems in fact that the outflow velocities are pretty comparable, and both velocities are high enough to trigger both fast and slow shocks in their sphere of influence. There is no fundamental difference from this aspect for the two galaxies in consideration.

**5.4.2 Cosmic-ray ionization rate (CRIR).** Generally the CRIR in the CMZ of NGC 253 is a few orders of magnitude higher than in the CND of NGC 1068. From Section 3 it was clear that the three shock-tracing species are all sensitive to CRIR, where high CRIR is suppressing the abundances of these species regardless of the gas density and shock velocity.

The assumed CRIR for the CND in NGC 1068 is based on the modeling conducted by Scourfield *et al.*[44] using multi-J CS observations. CS is often considered a dense gas tracer and could be more embedded in the gas clumps, therefore could be more shielded from radiation and cosmic rays. Indeed Scourfield *et al.*[44] showed that the gas densities traced by CS are generally at $n_{H_2} \sim 10^{5.5}$ to $10^7$ cm$^{-3}$, which are much higher than the densities traced by the CRIR tracer adopted for the NGC 253 study, *e.g.* Holdship *et al.*[47]. A follow-up study of $H_3O^+$ and SO in NGC 1068 should be done to see if the derived CRIR may in fact be different from the CS study.[44]

# 6 Conclusions

We have revisited the shock chemistry of two nearby galaxies, the AGN-host galaxy NGC 1068 and the starburst galaxy NGC 253, by analysing three shock tracers: SiO, HNCO, and $CH_3OH$. We presented a chemical modelling study of these two distinctively different environments and explored the potential of using HNCO and SiO in determining the shock structure and history, as well as in probing the origin(s) of shocks in the nuclei of the two galaxies. We also briefly discussed the potential of using $CH_3OH$ as an additional shock tracer. We briefly summarize here our main conclusions:

(1) SiO can only be enhanced by fast shocks, HNCO can only be enhanced by slow shocks, and $CH_3OH$ can respond to the presence of both slow and fast shocks.

(2) Generally a high CRIR suppresses the chemical abundances of all three shock tracers, especially for the highest CRIR case explored, $\zeta = 10^5\zeta_0$.

(3) In the lower CRIR regimes ($\zeta = 1$ to $10\zeta_0$), the abundance of HNCO and $CH_3OH$ can also be enhanced by thermal sublimation instead of shocks. This has to be taken into careful consideration when interpreting results especially from the CND of NGC 1068.

(4) There are more similarities than differences in the properties traced by HNCO and SiO in both systems. Follow-up studies, including the observations of more transitions of these species for NGC 1068, are needed to better constrain the gas properties of the shocked gas.

## Author contributions

KYH contributed to the conceptualization, chemical modeling and analysis, and article text discussion. SV contributed to the conceptualization, analysis, and article text discussion.

## Conflicts of interest

There are no conflicts to declare.

## Acknowledgements

This work is supported by European Research Council (ERC) Advanced Grant MOPPEX 833460.

## Notes and references

1 J. Bland-Hawthorn, J. F. Gallimore, L. J. Tacconi, E. Brinks, S. A. Baum, R. R. J. Antonucci and G. N. Cecil, *Astrophys. Space Sci.*, 1997, **248**, 9–19.
2 S. García-Burillo, F. Combes, A. Usero, S. Aalto, M. Krips, S. Viti, A. Alonso-Herrero, L. K. Hunt, E. Schinnerer, A. J. Baker, F. Boone, V. Casasola, L. Colina, F. Costagliola, A. Eckart, A. Fuente, C. Henkel, A. Labiano, S. Martín, I. Márquez, S. Muller, P. Planesas, C. Ramos Almeida, M. Spaans, L. J. Tacconi and P. P. van der Werf, *Astron. Astrophys.*, 2014, **567**, A125.
3 S. García-Burillo, F. Combes, C. Ramos Almeida, A. Usero, A. Alonso-Herrero, L. K. Hunt, D. Rouan, S. Aalto, M. Querejeta, S. Viti, P. P. van der Werf, H. Vives-Arias, A. Fuente, L. Colina, J. Martín-Pintado, C. Henkel, S. Martín, M. Krips, D. Gratadour, R. Neri and L. J. Tacconi, *Astron. Astrophys.*, 2019, **632**, A61.
4 S. Viti, S. García-Burillo, A. Fuente, L. K. Hunt, A. Usero, C. Henkel, A. Eckart, S. Martin, M. Spaans, S. Muller, F. Combes, M. Krips, E. Schinnerer, V. Casasola, F. Costagliola, I. Marquez, P. Planesas, P. P. van der Werf, S. Aalto, A. J. Baker, F. Boone and L. J. Tacconi, *Astron. Astrophys.*, 2014, **570**, A28.
5 A. McCormick, S. Veilleux and D. S. N. Rupke, *Astrophys. J.*, 2013, **774**, 126.
6 R. Rekola, M. G. Richer, M. L. McCall, M. J. Valtonen, J. K. Kotilainen and C. Flynn, *Mon. Not. R. Astron. Soc.*, 2005, **361**, 330–336.
7 A. K. Leroy, A. D. Bolatto, E. C. Ostriker, E. Rosolowsky, F. Walter, S. R. Warren, J. Donovan Meyer, J. Hodge, D. S. Meier, J. Ott, K. Sandstrom, A. Schruba, S. Veilleux and M. Zwaan, *Astrophys. J.*, 2015, **801**, 25.
8 G. J. Bendo, R. J. Beswick, M. J. D'Cruze, C. Dickinson, G. A. Fuller and T. W. B. Muxlow, *Mon. Not. R. Astron. Soc.: Lett.*, 2015, **450**, L80–L84.
9 K. Sakamoto, R.-Q. Mao, S. Matsushita, A. B. Peck, T. Sawada and M. C. Wiedner, *Astrophys. J.*, 2011, **735**, 19.
10 D. K. Strickland, T. M. Heckman, K. A. Weaver and M. Dahlem, *Astron. J.*, 2000, **120**, 2965–2974.
11 D. K. Strickland, T. M. Heckman, K. A. Weaver, C. G. Hoopes and M. Dahlem, *Astrophys. J.*, 2002, **568**, 689–716.
12 M. S. Westmoquette, L. J. Smith and I. Gallagher, *Mon. Not. R. Astron. Soc.*, 2011, **414**, 3719–3739.

13 B. E. Turner, *Astrophys. J.*, 1985, **299**, 312–333.
14 A. D. Bolatto, S. R. Warren, A. K. Leroy, F. Walter, S. Veilleux, E. C. Ostriker, J. Ott, M. Zwaan, D. B. Fisher, A. Weiss, E. Rosolowsky and J. Hodge, *Nature*, 2013, **499**, 450–453.
15 F. Walter, A. D. Bolatto, A. K. Leroy, S. Veilleux, S. R. Warren, J. Hodge, R. C. Levy, D. S. Meier, E. C. Ostriker, J. Ott, E. Rosolowsky, N. Scoville, A. Weiss, L. Zschaechner and M. Zwaan, *Astrophys. J.*, 2017, **835**, 265.
16 N. Krieger, A. D. Bolatto, F. Walter, A. K. Leroy, L. K. Zschaechner, D. S. Meier, J. Ott, A. Weiss, E. A. C. Mills, R. C. Levy, S. Veilleux and M. Gorski, *Astrophys. J.*, 2019, **881**, 43.
17 R. C. Levy, A. D. Bolatto, A. K. Leroy, M. C. Sormani, K. L. Emig, M. Gorski, L. Lenkić, E. A. C. Mills, E. Tarantino, P. Teuben, S. Veilleux and F. Walter, *Astrophys. J.*, 2022, **935**, 19.
18 P. J. McCarthy, W. van Breugel and T. Heckman, *Astron. J.*, 1987, **93**, 264.
19 F. Müller-Sánchez, O. González-Martín, J. A. Fernández-Ontiveros, J. A. Acosta-Pulido and M. A. Prieto, *Astrophys. J.*, 2010, **716**, 1166–1177.
20 B. D. Lehmer, D. R. Wik, A. E. Hornschemeier, A. Ptak, V. Antoniou, M. K. Argo, K. Bechtol, S. Boggs, F. E. Christensen, W. W. Craig, C. J. Hailey, F. A. Harrison, R. Krivonos, J. C. Leyder, T. J. Maccarone, D. Stern, T. Venters, A. Zezas and W. W. Zhang, *Astrophys. J.*, 2013, **771**, 134.
21 J. L. Turner and P. T. P. Ho, *Astrophys. J. Lett.*, 1985, **299**, L77–L81.
22 S. García-Burillo, J. Martín-Pintado, A. Fuente and R. Neri, *Astron. Astrophys.*, 2000, **355**, 499–511.
23 D. S. Meier, F. Walter, A. D. Bolatto, A. K. Leroy, J. Ott, E. Rosolowsky, S. Veilleux, S. R. Warren, A. Weiß, M. A. Zwaan and L. K. Zschaechner, *Astrophys. J.*, 2015, **801**, 63.
24 P. K. Humire, C. Henkel, A. Hernández-Gómez, S. Martín, J. Mangum, N. Harada, S. Muller, K. Sakamoto, K. Tanaka, Y. Yoshimura, K. Nakanishi, S. Mühle, R. Herrero-Illana, D. S. Meier, E. Caux, R. Aladro, R. Mauersberger, S. Viti, L. Colzi, V. M. Rivilla, M. Gorski, K. M. Menten, K. Y. Huang, S. Aalto, P. P. van der Werf and K. L. Emig, *Astron. Astrophys.*, 2022, **663**, A33.
25 N. Harada, S. Martín, J. G. Mangum, K. Sakamoto, S. Muller, V. M. Rivilla, C. Henkel, D. S. Meier, L. Colzi, M. Yamagishi, K. Tanaka, K. Nakanishi, R. Herrero-Illana, Y. Yoshimura, P. K. Humire, R. Aladro, P. P. van der Werf and K. L. Emig, *Astrophys. J.*, 2022, **938**, 80.
26 J. Martín-Pintado, P. de Vicente, A. Fuente and P. Planesas, *Astrophys. J. Lett.*, 1997, **482**, L45–L48.
27 S. Hüttemeister, G. Dahmen, R. Mauersberger, C. Henkel, T. L. Wilson and J. Martin-Pintado, *Astron. Astrophys.*, 1998, **334**, 646–658.
28 I. Zinchenko, C. Henkel and R. Q. Mao, *Astron. Astrophys.*, 2000, **361**, 1079–1094.
29 I. Jiménez-Serra, P. Caselli, J. Martín-Pintado and T. W. Hartquist, *Astron. Astrophys.*, 2008, **482**, 549–559.
30 S. Martín, M. A. Requena-Torres, J. Martín-Pintado and R. Mauersberger, *Astrophys. J.*, 2008, **678**, 245–254.
31 N. J. Rodríguez-Fernández, M. Tafalla, F. Gueth and R. Bachiller, *Astron. Astrophys.*, 2010, **516**, A98.
32 D. S. Meier and J. L. Turner, *Astrophys. J.*, 2005, **618**, 259–280.

33 A. Usero, S. García-Burillo, J. Martín-Pintado, A. Fuente and R. Neri, *Astron. Astrophys.*, 2006, **448**, 457–470.
34 S. Martín, J. Martín-Pintado and R. Mauersberger, *Astrophys. J.*, 2009, **694**, 610–617.
35 S. García-Burillo, A. Usero, A. Fuente, J. Martín-Pintado, F. Boone, S. Aalto, M. Krips, R. Neri, E. Schinnerer and L. J. Tacconi, *Astron. Astrophys.*, 2010, **519**, A2.
36 D. S. Meier and J. L. Turner, *Astrophys. J.*, 2012, **755**, 104.
37 S. Martín, K. Kohno, T. Izumi, M. Krips, D. S. Meier, R. Aladro, S. Matsushita, S. Takano, J. L. Turner, D. Espada, T. Nakajima, Y. Terashima, K. Fathi, P. Y. Hsieh, M. Imanishi, A. Lundgren, N. Nakai, E. Schinnerer, K. Sheth and T. Wiklind, *Astron. Astrophys.*, 2015, **573**, A116.
38 G. Kelly, S. Viti, S. García-Burillo, A. Fuente, A. Usero, M. Krips and R. Neri, *Astron. Astrophys.*, 2017, **597**, A11.
39 K. Y. Huang, S. Viti, J. Holdship, S. García-Burillo, K. Kohno, A. Taniguchi, S. Martín, R. Aladro, A. Fuente and M. Sánchez-García, *arXiv*, 2022, preprint, arXiv:2202.05005, DOI: 10.48550/arXiv.2202.05005.
40 K.-Y. Huang, S. Viti, J. Holdship, J. G. Magnum, S. Martín, N. Harada, S. Muller, K. Sakamoto, K. Tanaka, Y. Yoshimura, R. Herrero-Illana, D. S. Meier, E. Behrens, P. P. van der Werf, C. Henkel, S. García-Burillo, V. M. Rivilla, K. L. Emig, L. Colzi, P. K. Humire, R. Aladro and M. Bouvier, *arXiv*, 2023, preprint, arXiv:2303.12685, DOI: 10.48550/arXiv.2303.12685.
41 P. Schilke, C. M. Walmsley, G. Pineau des Forets and D. R. Flower, *Astron. Astrophys.*, 1997, **321**, 293–304.
42 G. Fedoseev, S. Ioppolo, D. Zhao, T. Lamberts and H. Linnartz, *Mon. Not. R. Astron. Soc.*, 2015, **446**, 439–448.
43 A. López-Sepulcre, A. A. Jaber, E. Mendoza, B. Lefloch, C. Ceccarelli, C. Vastel, R. Bachiller, J. Cernicharo, C. Codella, C. Kahane, M. Kama and M. Tafalla, *Mon. Not. R. Astron. Soc.*, 2015, **449**, 2438–2458.
44 M. Scourfield, S. Viti, S. García-Burillo, A. Saintonge, F. Combes, A. Fuente, C. Henkel, A. Alonso-Herrero, N. Harada, S. Takano, T. Nakajima, S. Martín, M. Krips, P. P. van der Werf, S. Aalto, A. Usero and K. Kohno, *Mon. Not. Roy. Astron. Soc.*, 2020, **496**, 5308–5329.
45 J. Holdship, S. Viti, S. Martín, N. Harada, J. Mangum, K. Sakamoto, S. Muller, K. Tanaka, Y. Yoshimura, K. Nakanishi, R. Herrero-Illana, S. Mühle, R. Aladro, L. Colzi, K. L. Emig, S. García-Burillo, C. Henkel, P. Humire, D. S. Meier, V. M. Rivilla and P. van der Werf, *Astron. Astrophys.*, 2021, **654**, A55.
46 N. Harada, S. Martín, J. G. Mangum, K. Sakamoto, S. Muller, K. Tanaka, K. Nakanishi, R. Herrero-Illana, Y. Yoshimura, S. Mühle, R. Aladro, L. Colzi, V. M. Rivilla, S. Aalto, E. Behrens, C. Henkel, J. Holdship, P. K. Humire, D. S. Meier, Y. Nishimura, P. P. van der Werf and S. Viti, *Astrophys. J.*, 2021, **923**, 24.
47 J. Holdship, J. G. Mangum, S. Viti, E. Behrens, N. Harada, S. Martín, K. Sakamoto, S. Muller, K. Tanaka, K. Nakanishi, R. Herrero-Illana, Y. Yoshimura, R. Aladro, L. Colzi, K. L. Emig, C. Henkel, Y. Nishimura, V. M. Rivilla and P. P. van der Werf, *arXiv*, 2022, preprint, arXiv:2204.03668, DOI: 10.48550/arXiv.2204.03668.
48 E. Behrens, J. G. Mangum, J. Holdship, S. Viti, N. Harada, S. Martin, K. Sakamoto, S. Muller, K. Tanaka, K. Nakanishi, R. Herrero-Illana,

Y. Yoshimura, R. Aladro, L. Colzi, K. L. Emig, C. Henkel, K.-Y. Huang, P. K. Humire, D. S. Meier and V. M. Rivilla, *arXiv*, 2022, preprint, arXiv:2209.06244, DOI: 10.48550/arXiv.2209.06244.
49 J. Holdship, S. Viti, I. Jiménez-Serra, A. Makrymallis and F. Priestley, *Astron. J.*, 2017, **154**, 38.
50 A. G. G. M. Tielens and W. Hagen, *Astron. Astrophys.*, 1982, **114**, 245–260.
51 S. B. Charnley, A. G. G. M. Tielens and T. J. Millar, *Astrophys. J. Lett.*, 1992, **399**, L71.
52 K. Hiraoka, N. Ohashi, Y. Kihara, K. Yamamoto, T. Sato and A. Yamashita, *Chem. Phys. Lett.*, 1994, **229**, 408–414.
53 N. Watanabe and A. Kouchi, *Astrophys. J. Lett.*, 2002, **571**, L173–L176.
54 G. W. Fuchs, H. M. Cuppen, S. Ioppolo, C. Romanzin, S. E. Bisschop, S. Andersson, E. F. van Dishoeck and H. Linnartz, *Astron. Astrophys.*, 2009, **505**, 629–639.
55 J. C. Santos, K.-J. Chuang, T. Lamberts, G. Fedoseev, S. Ioppolo and H. Linnartz, *Astrophys. J. Lett.*, 2022, **931**, L33.
56 W. D. Geppert, F. Hellberg, F. Österdahl, J. Semaniak, T. J. Millar, H. Roberts, R. D. Thomas, M. Hamberg, M. A. Ugglas, A. Ehlerding, V. Zhaunerchyk, M. Kaminska and M. Larsson, *Astrochemistry: Recent Successes and Current Challenges*, 2005, pp. 117–124.
57 R. Garrod, I. H. Park, P. Caselli and E. Herbst, *Faraday Discuss.*, 2006, **133**, 51.
58 L. Bizzocchi, P. Caselli, S. Spezzano and E. Leonardo, *Astron. Astrophys.*, 2014, **569**, A27.
59 C. Vastel, C. Ceccarelli, B. Lefloch and R. Bachiller, *Astrophys. J. Lett.*, 2014, **795**, L2.